# Gradient-based Representational Similarity Analysis with Searchlight for Analyzing fMRI Data

Xiaoliang Sheng，Muhammad Yousefnezhad，Tonglin Xu，Ning Yuan, and Daoqiang Zhang

College of Computer Science and Technology,
Nanjing University of Aeronautics and Astronautics, Nanjing 211106, China

dqzhang@nuaa.edu.cn

**Abstract.** Representational Similarity Analysis (RSA) aims to explore similarities between neural activities of different stimuli. Classical RSA techniques employ the inverse of the covariance matrix to explore a linear model between the neural activities and task events. However, calculating the inverse of a large-scale covariance matrix is time-consuming and can reduce the stability and robustness of the final analysis. Notably, it becomes severe when the number of samples is too large. For facing this shortcoming, this paper proposes a novel RSA method called gradient-based RSA (GRSA). Moreover, the proposed method is not restricted to a linear model. In fact, there is a growing interest in finding more effective ways of using multi-subject and whole-brain fMRI data. Searchlight technique can extend RSA from the localized brain regions to the whole-brain regions with smaller memory footprint in each process. Based on Searchlight, we propose a new method called Spatiotemporal Searchlight GRSA (SSL-GRSA) that generalizes our ROI-based GRSA algorithm to the whole-brain data. Further, our approach can handle some computational challenges while dealing with large-scale, multi-subject fMRI data. Experimental studies on multi-subject datasets confirm that both proposed approaches achieve superior performance to other state-of-the-art RSA algorithms.

## 1 Introduction

One of the most significant challenges in brain decoding is finding some more effective ways of using multi-subject and whole-brain fMRI data. Representational Similarity Analysis (RSA) is one of the fundamental approaches in fMRI analysis and evaluates similarities between different cognitive tasks [1-3]. Here, one subject is scanned while watching different visual stimuli. With different pairs of stimuli, the brain generates corresponding patterns of neural activities, and then the RSA calculates the similarities between the neural activity patterns of different stimuli. This process obtains Representational Similarity Matrix (RSM), and the matrix encodes the similarity structure. The goal of the method is to explore the correlation between dif-ferent cognitive tasks. Figure 1 shows the computation of the representational similarity matrix (RSM).

RSA can be casted as a multi-task regression problem. Classical RSA is based on basic linear approaches, e.g., Ordinary Least Squares (OLS) or General [1-2]. Indeed, these methods are restricted to a linear model, each data contains a large number of voxels, and the number of voxels far exceeds the time points. The methods mentioned cannot obtain satisfactory results on fMRI datasets. Moreover, the data is difficult to be converted into a matrix by this method [4], and it could reduce the stability and robustness of the final analysis when the Signal-to-Noise Ratio (SNR) is low [7].

For OLS and GLM, they face a problem of overfitting. The current approaches consider that the regularization can avoid overfitting. For example, Least Absolute Shrinkage and Selection Operator (LASSO) method employs norm $\ell 1$ to address the regression problem [9], whereas Ridge Regression method uses the norm $\ell 2$ to deal with the mentioned problem [8]. As an alternative approach, the Elastic Net method handle above issue by employing $\ell 1$ and $\ell 2$ norms [10].

In general, The RSA provides a way to compare different representational geometries across subjects, brain regions, measurement modalities, and even species. Since the similarity structure can be estimated from the imaging data even if the coding model is not constructed, RSA is suitable not only for model testing but also for exploratory research [3]. Indeed, RSA is initially used as a tool to study visual representations [2, 5-6], semantic representations [12-13], and lexical representations [14]. Further, RSA is utilized to reveal the network about dimensions of social-information representations [15-16].

As an alternative to region-of-interest based analysis, researchers introduce the 'searchlight' approach that performs multivariate analysis on sphere-shaped groups of voxels centered on each brain voxel one by one [1]. Nowadays, fMRI brain image datasets have a large number of subjects. Thus the whole-brain datasets are high-dimensional. In the current general RSA algorithm, the data is difficult to be converted into a matrix by this method and the inverse of the voxel matrix cannot be avoided. Besides, the optimization of RSA is difficult when the number of voxels is too large. Fortunately, modern RSA algorithm can optimize the solution process in comparison to traditional RSA method [17]. One of the modern RSA methods utilizes the searchlight technique, which is applied to EMEG [14]. As a novel application, the searchlight RSA method can be utilized to analyze the structure of moral violations space [11].

In this paper, we propose a new RSA method based on gradient descent called Gradient Representational Similarity Analysis (GRSA). The Gradient RSA algorithm can handle the RSA problem by calculating the solution of LASSO using stochastic gradient descent. It can solve the mapping feature matrix by using stochastic gradient descent method with iteration to obtain an optimal result and explore the similarity between different neural activity patterns. Another key contribution of this paper is a novel application for Searchlight. GRSA is a tool for analyzing whether localized brain regions encode cognitive similarities. Using searchlight, we propose a new method called spatiotemporal searchlight GRSA (SSL-GRSA). In chapter 3.2, we

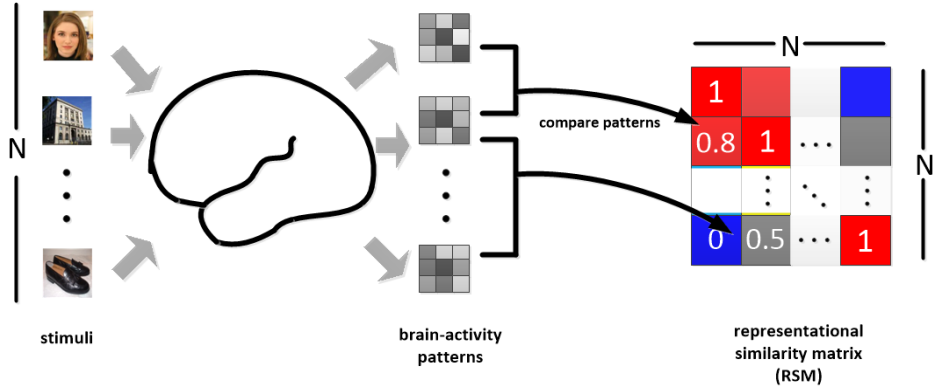

**stimuli**

**brain-activity patterns**

**representational similarity matrix (RSM)**

**Fig. 1.** Computation of the representational similarity matrix (RSM). The matrix encodes the similarity structure. Each block in the RSM is a correlation distance between activation patterns of a pair of experimental conditions (or stimuli). The elements on the main diagonal of the matrix are one by definition. In the non-diagonal part of RSM, a larger value indicates that two stimuli have a high similarity, and the small value implies that the two stimuli are not similar.

focus on this approach with an aim to link searchlight analysis with GRSA. We develop this model by using a spatiotemporal searchlight GRSA algorithm which can generalize our ROI-based GRSA algorithm to the whole-brain data.

## 2 Representational Similarity Analysis (RSA)

The application of RSA is based on a general linear model (GLM). This method assumes that the neural pattern of fMRI responses is related to stimuli events.

$$\boldsymbol{Y}^{(\ell)} = \boldsymbol{X}^{(\ell)}\boldsymbol{B}^{(\ell)} + \boldsymbol{\epsilon}^{(\ell)} \tag{1}$$

where $\boldsymbol{Y}^{(\ell)} = \left\{ y_{ij} \right\} \in \mathbb{R}^{T \times V}, 1 \leq i \leq T, 1 \leq j \leq V$ denotes the fMRI time series from $\ell$-th subject, $T$ is the number of time points and $V$ is the number of brain voxels. Design matrix is denoted by $\boldsymbol{X}^{(\ell)} = \{x_{ik}\} \in \mathbb{R}^{T \times P}, 1 \leq i \leq T, 1 \leq k \leq P$. The design matrix is obtained by the convolution of the time series of the stimuli with a typical hemodynamic response function (HRF). Here, $P$ denotes the number of distinct categories of stimuli, $\boldsymbol{B}^{(\ell)} = \left\{ \beta_{kj} \right\} \in \mathbb{R}^{P \times V}, \beta_{kj} \in \mathbb{R}, 1 \leq k \leq P, 1 \leq j \leq V$ denotes the matrix of estimated regressors, and $\beta_{kj}$ is an amplitude reflecting the response of $j$-th voxel to the $k$-th stimulus. This paper assumes that the neural activities of each subject are column-wise standardized, i.e., $\boldsymbol{Y}^{(\ell)} \sim \mathcal{N}(0,1)$. Indeed, RSA method is looking for the following objective function:

$$\min_{\boldsymbol{B}^{(\ell)}} \left\| \boldsymbol{Y}^{(\ell)} - \boldsymbol{X}^{(\ell)}\boldsymbol{B}^{(\ell)} \right\|_F^2 - r(\boldsymbol{B}^{(\ell)}) \tag{2}$$

where $r(\boldsymbol{B}^{(\ell)})$ is the regularization term for $\ell$-th subject. Notably, the regularization term is zero ($r(\boldsymbol{B}^{(\ell)}) = 0$) for non-regularized methods, including OLS and GLM. The term $r(\boldsymbol{B}^{(\ell)})$ is $\alpha\|\boldsymbol{B}\|_F^2$ for Ridge Regression, $\alpha\|\boldsymbol{B}\|_1$ for LASSO method, $\alpha\rho\|\boldsymbol{B}\|_1 + \frac{\alpha(1-\rho)}{2}\|\boldsymbol{B}\|_F^2$ for Elastic Net method.

In order to generalize RSA for multi-subject fMRI datasets, we calculate the mean of the regressors matrices across subjects:

$$\boldsymbol{B}^* = \frac{1}{S}\sum_{\ell=1}^{S}\boldsymbol{B}^{(\ell)} \tag{3}$$

where $S$ denotes the number of subjects, and each row of $\boldsymbol{B}^* \in \mathbb{R}^{P \times V} = \{\boldsymbol{\beta}_{1.}^*, \ldots, \boldsymbol{\beta}_{P.}^*\}$, $\boldsymbol{\beta}_{k.}^* \in \mathbb{R}^V$ illustrates the extracted neural signature belonging to $k$-th category of cognitive tasks.

Three metrics will be used to evaluate the performance of RSA methods. As the first metric, we calculate the mean of square error for analyzing the accuracy of regression:

$$MSE = \frac{1}{TSV}\sum_{\ell=1}^{S}\sum_{i=1}^{T}\sum_{j=1}^{V}\left(x_{ij}^{(\ell)} - \sum_{k=1}^{P}d_{ik}^{(\ell)}\beta_{kj}^{(\ell)}\right)^2 \tag{4}$$

The next two techniques evaluate between-class correlation and between-class covariance of the regressors matrices:

$$CR = \frac{1}{S}\sum_{\ell=1}^{S}\max_{\substack{1 \le i \le P \\ i < j \le P}}\left\{Corr(\beta_{i.}^{(\ell)}, \beta_{j.}^{(\ell)})\right\} \tag{5}$$

$$CV = \frac{1}{S}\sum_{\ell=1}^{S}\max_{\substack{1 \le i \le P \\ i < j \le P}}\left\{Cov(\beta_{i.}^{(\ell)}, \beta_{j.}^{(\ell)})\right\} \tag{6}$$

where $\beta_{i.}^{(\ell)}, \beta_{j.}^{(\ell)}$ are rows of $\boldsymbol{B}^{(\ell)}$, function $Corr$ is the Pearson correlation, and function $Cov$ calculates the covariance between two vectors. All of these three metrics must be minimized for an ideal solution [7, 17].

## 3    Gradient Representational Similarity Analysis (GRSA)

fMRI brain data is high-dimensional. In fMRI, each data contains a large number of voxels, and the number of voxels far exceeds the time points. Meanwhile, the presence of similarity of different features leads to some redundant information. Feature selection can solve this problem. Therefore, we use the $\ell1$ norm here. The objective function is optimized as follows:

$$J(\boldsymbol{B}^{(\ell)}) = \min_{\boldsymbol{B}^{(\ell)}} \ L(\boldsymbol{B}^{(\ell)}) + r(\boldsymbol{B}^{(\ell)}) \tag{7}$$

where the typical loss functions considered here are squared Frobenius error, i.e., $L(\boldsymbol{B}^{(\ell)}) = \left\|\boldsymbol{Y}^{(\ell)} - \boldsymbol{X}^{(\ell)} \cdot \boldsymbol{B}^{(\ell)}\right\|_F^2$, and $r(\boldsymbol{B}^{(\ell)})$ is the $\ell1$ norm defined as $\alpha\|\boldsymbol{B}\|_1$. The problem of this approach is that the computation complexity is tremendous when there are a large number of features. And this method is merely applies to the linear model.

### 3.1 Optimization

In this section, we attempt to propose a method that is not restricted to a linear model and can reduce the time complexity on high-dimensional data. Here, we propose an effective approach that utilizes Stochastic Gradient Descent (SGD) for optimizing the LASSO objective function. In order to efficiently optimize (7), one solution is to calculate the gradient of (7) which is needed in Stochastic Gradient Descent (SGD) algorithm. The step of gradient optimization is as follows:

$$\nabla J\big(\boldsymbol{B}_t^{(\ell)}\big) = \frac{\partial}{\partial \boldsymbol{B}_t^{(\ell)}} J\big(\boldsymbol{B}^{(\ell)}\big) \tag{8}$$

$$\boldsymbol{B}_{t+1}^{(\ell)} = \boldsymbol{B}_t^{(\ell)} - \alpha^t \nabla J\big(\boldsymbol{B}_t^{(\ell)}\big)$$

where $\nabla J\big(\boldsymbol{B}_t^{(\ell)}\big)$ denotes the gradient of $J\big(\boldsymbol{B}^{(\ell)}\big)$ from $t$-th iteration. The step of iteration of $\boldsymbol{B}^{(\ell)}$ denoted as (9). $\alpha^t$ is the self-adaptive learning rate ,which is defined as follows:

$$\alpha^t = \frac{\alpha}{\sqrt{t+1}} \tag{10}$$

Here, $t \in \mathbb{R}$ is the number of iterations. $\alpha^t$ denotes the updated learning rate of $t$-th iteration. Since different features have different ranges of values, the iteration could be very slow. In order to apply this algorithm to fMRI brain datasets, the SGD algorithm randomly selects a batch of the time points instead of the whole time points to update the model parameters. So each time of learning is fast and the model parameters can be updated online. This paper uses GRSA approach for estimating the optimized solution. GRSA can reduce the time complexity when applied to fMRI brain datasets, and explore the similarity between different neural activity patterns by iterative optimal algorithm. Our method can rapidly reduce the time complexity and have smaller memory footprint in each process. This application of GRSA could be used not only in the linear model but also in the non-linear model.

### 3.2 Spatiotemporal Searchlight GRSA (SSL-GRSA)

Finding the most effective method for analyzing multi-subject fMRI data is a long-standing and challenging problem. Since the scarcity of data for each subject and the differences of brain anatomy and functional response between different subjects, researchers have an increasing interest in human cognitive fMRI research. Multi-subject fMRI datasets contain two group datasets, i.e., Region of Interests (ROI) based datasets, and whole-brain datasets. The ROI-based method analyzes the representation structure in a set of predefined brain regions. However, other brain regions also have representational structures that are suitable for the prediction of our model. Whole-brain data can be used to figure out what information is represented in a region of the human brain. People want to find some more effective ways to analyze whole-brain data. Searchlight analysis provides a way to map cube-shaped groups of voxels across the whole brain continuously [1]. Therefore, we propose a method that combines the ideas of the GRSA model and searchlight-based technique to analyze multi-subject whole-brain fMRI data. A searchlight version of GRSA is conceptually new. Therefore, we refer to our method as Searchlight GRSA (SSL-GRSA).

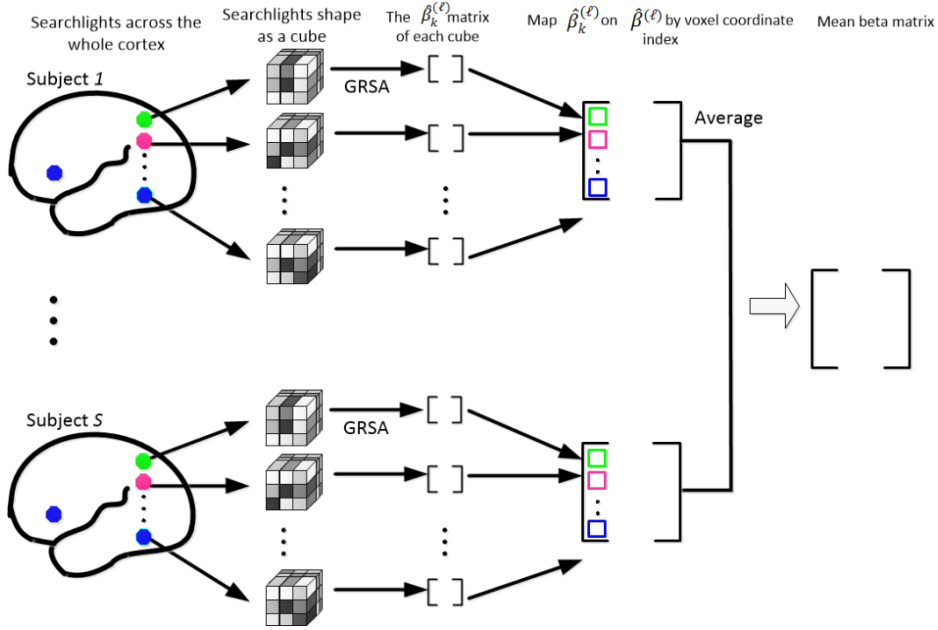

**Fig. 2.** Process of Spatiotemporal Searchlight GRSA (SL-GRSA). The whole-brain data of each subject is divided into K cubes (searchlights) with a specified size. Here, this size is fixed as $3 \times 3 \times 3$. Then, the GRSA approach applies to each cube to generate K local matrices denoted by $\hat{\beta}_k^{(\ell)}$. In the end, we splice those K local $\hat{\beta}_k^{(\ell)}$ matrices into a complete $\hat{\beta}^{(\ell)}$ matrix according to the coordinates of voxels. The mean matrix is obtained by averaging over all matrices $\hat{\beta}^{(\ell)}$.

$\hat{Y}^{(\ell)} \in \mathbb{R}^{v_x \times v_y \times v_z \times T}$ of four dimension is fMRI time series data from $\ell$-th subject where $1 \leq \ell \leq S$ and $S$ is the number of subjects. The tuple $(x, y, z)$ refers to the standard axes, whereas $v_x, v_y, v_z$ refer to the number of voxels along the corresponding axis respectively, and $T$ is the number of time samples in units of repetition time (TR). The process of our searchlight method is as follows: Firstly, a sliding cube is selected and the cube at a specific time covers a contiguous region of voxels. The selected snapshots of the cube need to be adjacent and avoid overlapping. Then,the voxels of the whole-brain is then analyzed by spatial local analysis in each cube. GRSA method is applied to cube groups of voxels in a line. Therefore, the ROI method can be extended to the whole-brain data. The process of our method is depicted in Figure 2.

**Table 1.** The datasets.

| Title | ID | Task Type | S | P | T | Scan | TR | TE |
|---|---|---|---|---|---|---|---|---|
| Visual object recognition | R105 | visual | 6 | 8 | 121 | G3T | 2500 | 30 |
| Word and object processing | R107 | visual | 49 | 4 | 164 | S3T | 2000 | 28 |
| Weather prediction without feed-back | W011 | decision | 14 | 4 | 236 | S3T | 2000 | 25 |
| Selective stop signal task | W017 | decision | 8 | 6 | 546 | S3T | 2000 | 25 |
| Weather prediction | W052 | decision | 13 | 2 | 450 | S3T | 2000 | 20 |

This paper utilizes five datasets, shared by Open fMRI (http://openfmri.org). S is the number of subject, P denotes the number of stimulus categories, T is the number of scans in unites of scans in unites of TRs (Time of Repetition), $V_{ROI}$ denotes the number of voxels in ROI. In the column of Scan, G = General Electric, or S = Siemens in 3 Tesla. TR is Time of Repetition in millisecond and TE denotes Echo Time in millisecond.

For standard Searchlight-based RSA method, the study first used the scene image as task stimuli for experiment, and then used the Searchlight method to find brain regions related to the perception of human brain. The results show that using the searchlight method, we can find the active brain regions in the FMRI data related to scene recognition of each subject. Compared with standard searchlight RSA, our method is competitive and performs better with the same cube size. It's worth mentioning that we only load necessary data according to the mini batch to maintain a reduced memory footprint in each process. We extend the application of GRSA from ROI to the whole-brain. Further, we create a novel approach that addresses some computational challenges while dealing with large-scale, multi-subject fMRI data.

## 4 Experiments

### 4.1 Datasets

This paper utilizes five datasets, shared by Open fMRI (http://openfmri.org), for running empirical studies. All datasets are separately preprocessed by FSL 5.0.10 (https://fsl.fmrib.ox.ac.uk), i.e., slice timing, anatomical alignment, normalization, smoothing. Here, we use two groups of datasets, i.e., Region of Interests (ROI) based datasets, and whole-brain datasets. Here, we analyze some specific parts of brain images in ROI-based data, where these parts are manually selected based on the original papers of each data. In this paper, we use 'R' prefix for the ROI-based dataset and a 'W' prefix is used for denoting the whole-brain data.

Technically, the whole-brain datasets include all of the neural activities which are registered to a standard space, i.e., Montreal Neurological Institute (MNI) 152 space $T1$ with voxel size 4mm. Before applying our approach to each fMRI dataset, the dataset is normalized, i.e., $Y^{(\ell)} \sim \mathcal{N}(0,1)$, which allows us to obtain desirable experiment result. The technical information of these datasets is shown in Table 1.

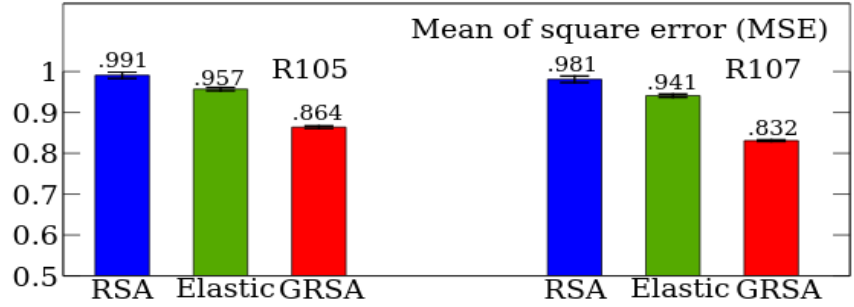

**Fig. 3.** The standard deviation of MSE for all RSA methods in the Figure 3 is lower than $10^{-2}$.

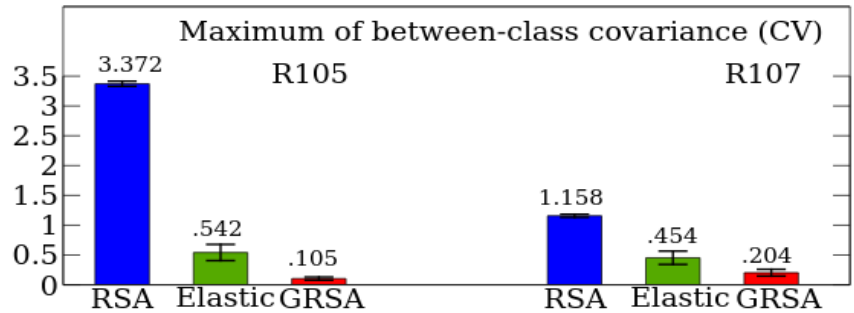

**Fig. 4.** Maximum of between-class covariance (*CV*) across subjects.

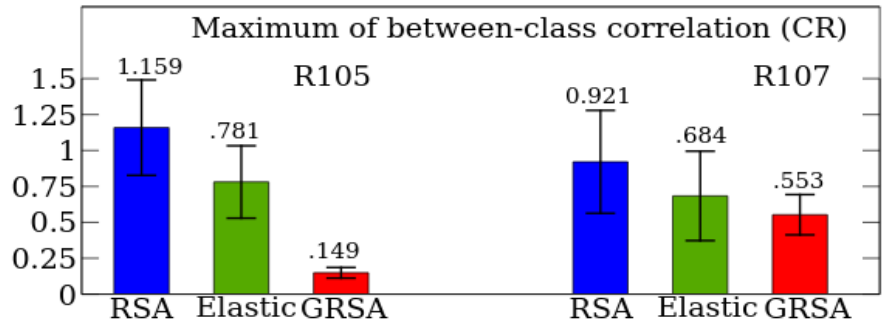

**Fig. 5.** Maximum of between-class correlation (*CR*) across subjects.

## 4.2 ROI data analysis

In this section, we analyze the performance of our method results by calculating three metrics, including mean of square error (MSE), the maximum of between-class covariance (CV), and the maximum of between-class correlation (CR).We use the ROI data in each experiment, thus R105 dataset and R107 dataset are selected from five differ-ent datasets. In order to create the comparative experiments, we use the classical RSA based on GLM as a baseline. Elastic Net is employed for the empirical research. In this method, the best results are obtained when the parameters are $\alpha = 1.0$ and $\rho = 0.5$. Moreover, GRSA generates the results by setting $\alpha = 0.9$. The number of iterations for our method is considered 1000. The batch size is set 50 and learning rates of normalized datasets is $10^{-3}$ .

Figure 3 shows the test results of MSE, which is non-negative. MSE is an indicator used to reflect the quality of the estimator. the smaller the MSE is, the better the method is. Further, MSE is calculated by Formula (4). The results of our method in comparison to other methods are shown in Figure 3. GRSA has the best results compared to other RSA methods. The standard deviation of MSE for all RSA methods in the Table 2 is lower than $10^{-2}$.

Figure 4 has analyzed the maximum of between-class covariance by using (6). The maximum of between-class covariance can be calculated as the maximum value ranging over all different pairs of stimuli. Moreover, Figure 5 has evaluated the maximum of between-class correlation by employing (5) in which it searches the maximum Pearson correlation coefficient amongst different pairs of stimuli. For those indicators, the smaller they are, the better the method analyzes the similarity between different neural activity patterns. Compared with other RSA methods in Figure 4 or Figure 5, GRSA has the best results.

### 4.3 Whole-brain data analysis

ROI is a manually selected area based on anatomical images of the brain. We analyze the potential information of the data through the ROI based method. However, a certain type of information is not necessarily confined to only one specific brain region, and could be included in several areas. Therefore, the analysis of the whole-brain data becomes more important. The GRSA method is applied to whole-brain data and this approach can explore the relationship between different cognitive tasks. In this paper, the whole-brain datasets are used in our method, i.e., W011 dataset, W017 dataset and W052 dataset.

In this section, we implement the comparative experiments by some traditional methods. We use the ordinary Spatiotemporal Searchlight RSA (SSL-RSA) as the baseline. For the empirical study, Spatiotemporal Searchlight Elastic Net (SSL- Elastic Net) is utilized. As mentioned before, both SSL-RSA and RSA share the same parameters. And so do SSL- Elastic Net and Elastic Net. Previously mentioned, the main challenges are the high dimension of data and the issue of memory footprint.

Our approach can address these challenges and has good performance. The cube size can be set arbitrarily. Thus, all Searchlight RSA methods take the same cube size set as $3 \times 3 \times 3$. In fact, the best result is obtained by using this cube size. The result of each contrast experiment is showed in Table 2 Table 3.

In each comparative experiment, we evaluate all the methods by using CV and CR. The formulas of these two indicators have already been mentioned in the previous section. Table 2 has analyzed the maximum of between-class covariance whereas

**Table 2.** Maximum of between-class covariance (*CV*) across subjects (max±std)

| Datasets | SSL-RSA | SSL-Elastic Net | SSL-GRSA |
|----------|---------|-----------------|----------|
| W011 | 0.415±0.125 | 0.265±0.046 | **0.208±0.042** |
| W017 | 0.462±0.062 | 0.237±0.186 | **0.143±0.143** |
| W052 | 1.831±0.184 | 0.396±0.143 | **0.237±0.052** |

**Table 3.** Maximum of between-class correlation (*CR*) across subjects (max±std)

| Datasets | SSL-RSA | SSL-Elastic Net | SSL-GRSA |
|----------|---------|-----------------|----------|
| W011 | 0.785±0.033 | **0.507±0.042** | 0.609±0.202 |
| W017 | 0.849±0.124 | 0.441±0.052 | **0.358±0.082** |
| W052 | 0.866±0.071 | 0.471±0.104 | **0.407±0.151** |

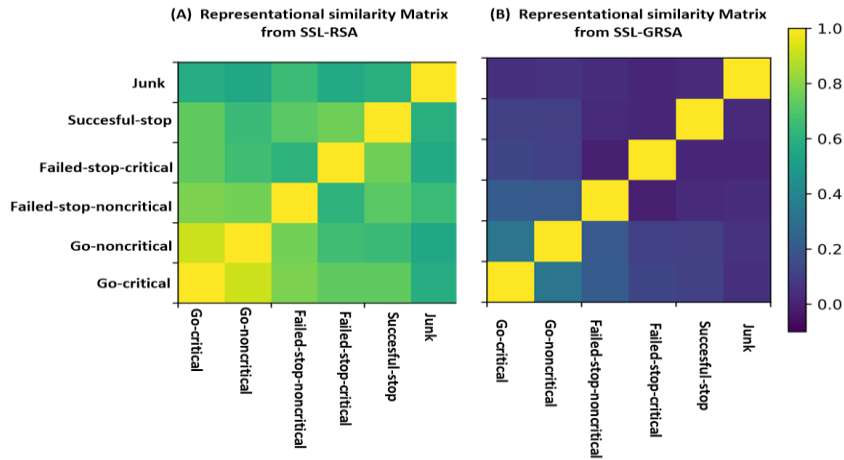

**Fig. 6.** Comparing correlation of a traditional method and SSL-GRSA method by using W017

Table 3 evaluated the maximum of between-class correlation. As depicted in the result Table 2, SSL-GRSA has generated better performance in comparison with other methods. Further, as Table 3 demonstrates, the performance of the maximum of between-class correlation is significantly lower except for W011, which confirms that our method is better.

Base on W017 data, Figure 6 depicts the comparison of correlation of a traditional method and SSL-GRSA method. Each small block shows the similarity of the related category of stimuli with respect to the corresponding row and column. Therefore, we compare the between-class correlation of SSL-GRSA with the traditional methods. SSL-GRSA provides the best similarity analysis compared with other methods.

### 4.4 Runtime analysis

This section analyzes the runtime of the proposed method and compares it to the runtime of other RSA methods. Here, the analysis is based on the ROI datasets. For convenience, the runtime of other methods is scaled based on GRSA, that is, the runtime of GRSA is regarded as a unit. As illustrated in Figure 7, the Elastic Net is the slowest one whereas traditional RSA beats others. Since GRSA utilizes a min-batch of time-points, it runs faster than the regularized method. As a conclusion, the performance of GRSA is more efficient. It is worth mentioning that the runtime of the whole brain dataset has the same tendency.

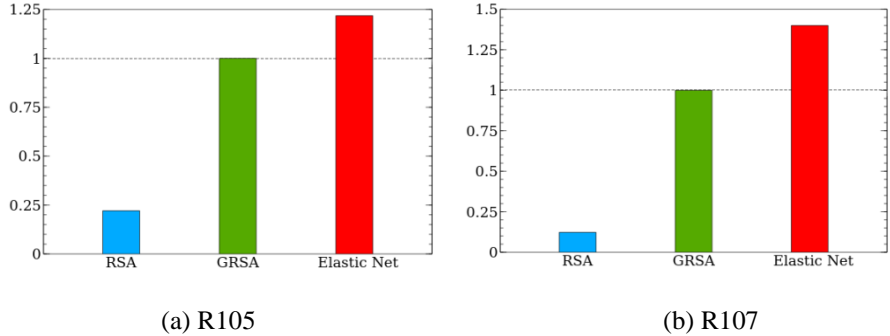

(a) R105           (b) R107

**Fig. 7.** Runtime Analysis

## 5 Conclusion

In this paper, we explored the method of Representational Similarity Analysis. we propose a novel RSA method called Gradient descent RSA. The Gradient-RSA algorithm handles the RSA problem by calculating the solution of LASSO using stochastic gradient descent, which is novel to RSA study. For the whole-brain data, the primary challenges are the high dimension of data and the issue of memory footprint. Another primary contribution of this paper is a new application in Searchlight. Based on Searchlight, the application of our GRSA method is extended from the localized brain regions to the whole-brain region. Further, Our methods show improved results over standard competing methods. In the future work, our method can be applied to more large-scale, multi-subject fMRI datasets, and further optimized by other new approaches to obtain better performance.

**Acknowledgements.** This work was supported in part by the National Natural Science Foundation of China under Grant (61876082, 61861130366, 61703301, and 61473149), the Fundamental Research Funds for the Central Universities and the Foundation of Graduate Innovation Center in NUAA (kfjj20171609).